\documentclass[10pt,twocolumn,english,prl,amsfonts,english,showpacs,reprint,floatfix]{revtex4}
\usepackage[T1]{fontenc}
\usepackage[latin9]{inputenc}
\usepackage{xcolor}
\usepackage{units}
\usepackage{amsmath}
\usepackage{amssymb}
\usepackage{graphicx}
\usepackage{esint}
\PassOptionsToPackage{normalem}{ulem}
\usepackage{ulem}

\makeatletter

\providecolor{lyxadded}{rgb}{0,0,1}
\providecolor{lyxdeleted}{rgb}{1,0,0}

\@ifundefined{textcolor}{}
{%
 \definecolor{BLACK}{gray}{0}
 \definecolor{WHITE}{gray}{1}
 \definecolor{RED}{rgb}{1,0,0}
 \definecolor{GREEN}{rgb}{0,1,0}
 \definecolor{BLUE}{rgb}{0,0,1}
 \definecolor{CYAN}{cmyk}{1,0,0,0}
 \definecolor{MAGENTA}{cmyk}{0,1,0,0}
 \definecolor{YELLOW}{cmyk}{0,0,1,0}
 }

\usepackage{babel}

\makeatother

\usepackage{babel}
\begin{document}
\global\long\def\graph{\Gamma}

\global\long\def\ps{p_{\sigma}}

\global\long\def\T{\mathbb{T}}

\global\long\def\R{\mathbb{R}}

\global\long\def\id{\mathbf{1}}

\global\long\def\Z{\mathbb{Z}}

\global\long\def\bL{\mathbf{L}}

\global\long\def\bA{\mathbf{A}}

\global\long\def\bS{\mathbf{S}}

\global\long\def\rmd{\mathrm{d}}

\global\long\def\rme{\mathrm{e}}

\global\long\def\diag{\operatorname{diag}}

\global\long\def\vol{\operatorname{vol}}

\global\long\def\spn{\operatorname{span}}

\global\long\def\mod{\operatorname{mod}}

\title{Universality of the momentum band density of periodic networks}

\author{Ram Band$^{1}$ and Gregory Berkolaiko$^{2}$}

\affiliation{$^{1}$School of Mathematics, University of Bristol, Bristol BS8
1TW, UK}

\affiliation{$^{2}$Department of Mathematics, Texas A\&M University, College
Station, TX 77843-3368, USA}
\begin{abstract}
The momentum spectrum of a periodic network (quantum graph) has a
band-gap structure. We investigate the relative density of the bands
or, equivalently, the probability that a randomly chosen momentum
belongs to the spectrum of the periodic network. We show that this
probability exhibits universal properties. More precisely, the probability
to be in the spectrum does not depend on the edge lengths (as long
as they are generic) and is also invariant within some classes of
graph topologies. 
\end{abstract}


\maketitle
The spectrum of Schr\"odinger operator in periodic medium is calculated
using the Floquet--Bloch procedure \citep{Kittel_solid63}: the periodic
medium is replaced with its fundamental domain endowed with parameter-dependent
quasi-periodic boundary conditions. The resulting parameter-dependent
spectrum is called the dispersion relation, and the range of the dispersion
relation is precisely the spectrum of the original structure. The
spectrum has a band-gap structure and knowing the band location and
sizes is of utmost importance in the theories of condensed matter
and of dielectric and acoustic media \citep{Yab_prl87,ZaaSawAll_prl85,FigKle_SIAM98,Luo+_s03,Kon+_apl10}.
Of particular recent interest is understanding the spectrum of quantum
graphs \citep{BK_graphs,GnuSmi_aip06}, motivated by their application
to solid state \citep{AvrExnLas_prl94,AkkComDebMonTex_ap00}, photonic
crystals \citep{KucKun_acm02}, carbon nano-structures \citep{KucPos_cmp07}
as well as their use as models for quantum chaos, both in theoretical
\citep{KotSmi_prl97,BerSchWhi_prl02,GnuAlt_prl04,GnuKeaPio_prl08,GnuSchSmi_prl13,JoyMulSie_arxiv13}
and experimental \citep{HulBauPak+_pre04,HulLawBau+_prl12} studies.

\begin{figure}[t]
\centering \includegraphics{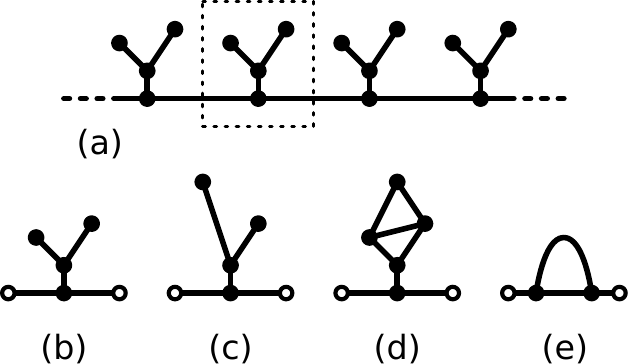} \caption{(a) An example of a $\Z^{1}$-periodic graph and (b) its fundamental
cell; (c)-(e) are other examples of the fundamental cell.}

\label{fig:per_1d} 
\end{figure}

In the present Letter we explore the \emph{relative} size of bands
and gaps and discover a curious universality. To be more precise,
we ask the following question: what is the probability $p_{s}$ that
a randomly and uniformly chosen momentum belongs to the spectrum of
the graph? For example, consider the $\Z^{1}$-periodic graphs of
Fig.~\ref{fig:per_1d}. How does $\ps$ change if we change the lengths
in the fundamental cell of the graph, from Fig.~\ref{fig:per_1d}(b)
to Fig.~\ref{fig:per_1d}(c)? How does $\ps$ change if we change
the topological structure to Fig.~\ref{fig:per_1d}(d) or \ref{fig:per_1d}(e)?

\begin{figure}
\includegraphics[scale=0.24]{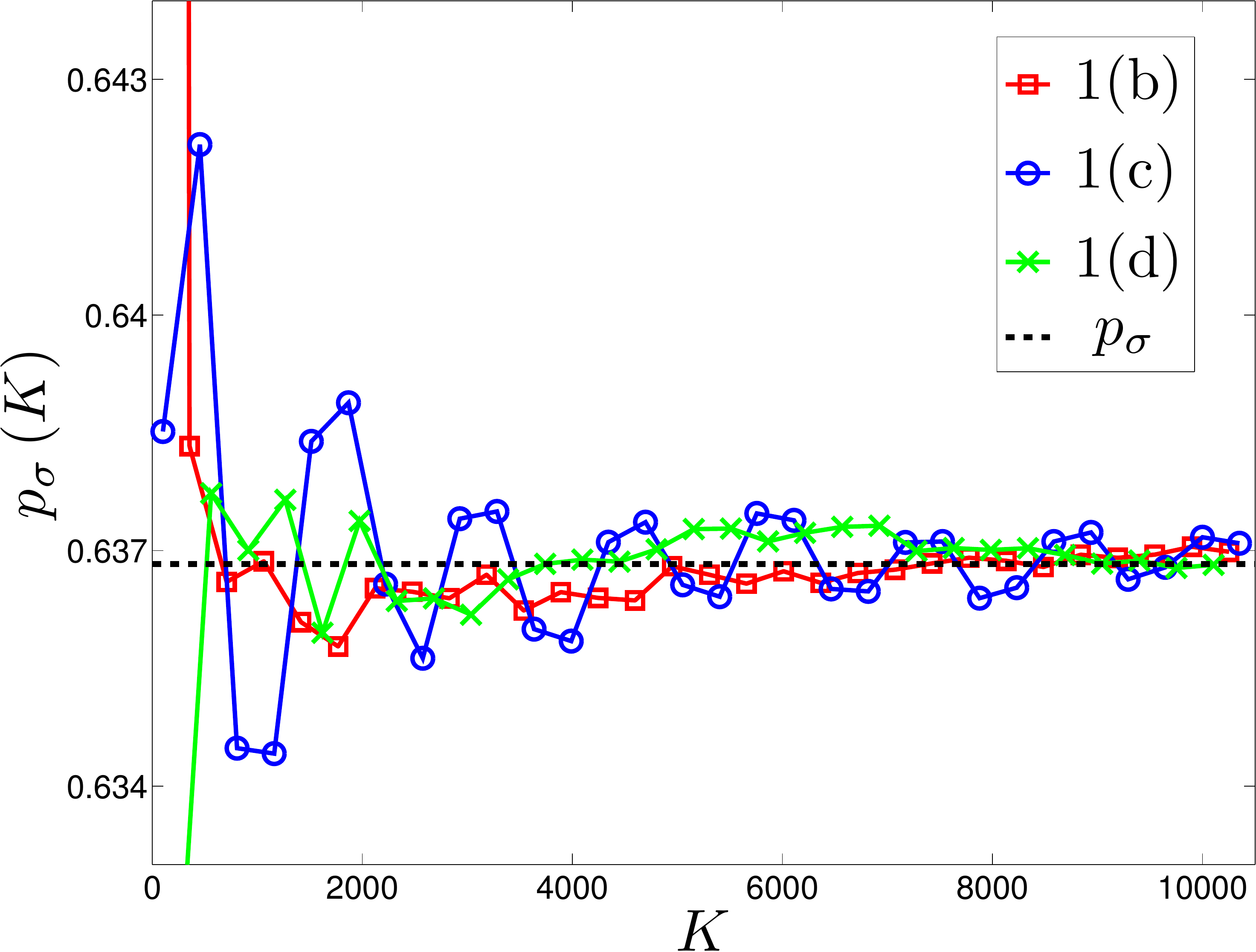} \caption{(color online) Numerical simulation of the convergence of $\ps\left(K\right)$
is shown for the three periodic graphs from Fig. 1(b)-(d). The limiting
value, $p_{\sigma}$, is shown as a dashed line. The graph lengths
are normalized such that $K$ equals the average number of spectral
bands.}

\label{fig:numerics} 
\end{figure}

Denote by $\ps\left(K\right)$ the probability of a uniformly chosen
momentum $k\in\left[0,K\right]$ to be in the spectrum and let $\ps:=\lim_{K\rightarrow\infty}\ps\left(K\right)$.
We find that the probability $\ps$ is well-defined and is \emph{independent}
of many features of the fundamental cell. In particular, all choices
in Fig.~\ref{fig:per_1d}(b) to (d) lead to the same value of $p_{\sigma}$
(assuming a generic choice of edge lengths). This is illustrated by
a numerical simulation in Fig. \ref{fig:numerics}. We will derive
the limiting value analytically below. Note that the value of $\ps$
for the cell in Fig.~\ref{fig:per_1d}(e) turns out to be different
from the others and will also be calculated.

Let us put the discussion onto a more formal footing. We consider
a $\Z^{d}$-periodic network of quantum wires on which we are solving
the spectral problem 
\begin{equation}
-\frac{d^{2}\psi}{dx^{2}}=k^{2}\psi,\label{eq:Hamiltonian}
\end{equation}
subject to the Kirchhoff--Neumann vertex conditions 
\begin{equation}
\begin{cases}
\psi(x)\text{ is continuous at }v,\\[0.5em]
\sum_{e\in\mathcal{E}_{v}}\frac{d\psi}{dx_{e}}(v)=0,
\end{cases}\label{eq:neumann_cond}
\end{equation}
where the sum is over the edges $\mathcal{E}_{v}$ emanating from
the vertex $v$ and the derivatives are taken into the edge. We denote
by $\sigma$ the set of $k$ values for which there is a solution
to (\ref{eq:Hamiltonian})-(\ref{eq:neumann_cond}); this is the momentum
spectrum of the graph. Now, the definition of $\ps$ can be formally
written as 
\begin{equation}
\ps=\lim_{K\to\infty}\ps\left(K\right)=\lim_{K\to\infty}\frac{1}{K}\Big|\sigma\cap[0,K]\Big|.\label{eq:p_s_probability}
\end{equation}

In this Letter we establish several properties of the probability
$\ps$. First of all, the above limit always exists. In addition,
if there is at least one gap in the spectrum, there are infinitely
many gaps and $\ps<1$. Similarly, if there is at least one non-flat
band, there are infinitely many and $\ps>0$. Finally, and perhaps
most strikingly, provided the lengths of edges in the fundamental
set are generic, the value of $\ps$ is \emph{independent} of their
precise value. We also find that the value of $\ps$ is independent
of some details of the cell's topology.


\paragraph{\uline{Secular equation and dispersion relation.}}

\begin{figure}[!]
  \setlength{\unitlength}{\columnwidth} 
  \begin{picture}(1,0.38) 
    \put(0.05,0.07){\includegraphics{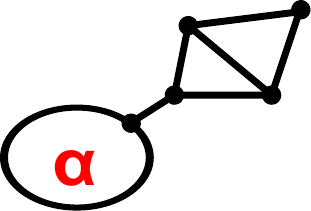}}
    \put(0.12,0){(a)}
    \put(0.55,0.07){\includegraphics{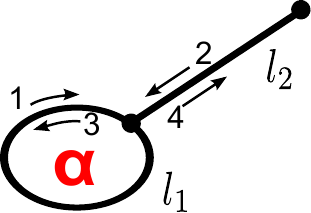}}
    \put(0.62,0){(b)}
  \end{picture}%
  \caption{(a) A graph consisting of a loop pierced by a magnetic flux and a
    decoration (b) Similar graph, but with a single edge decoration}
  \label{fig:loop_and_dangling_graph} 
\end{figure}

In the Floquet--Bloch procedure for quantum graphs (see, e.g., \citep{BK_graphs})
we identify a set of $J$ generators of the lattice of periods and
assign to each a quasi-momentum variable $\alpha_{j}$, $j=1,\ldots,J$.
If the vertices $v_{+}$ and $v_{-}$ of the fundamental cell are
identified by the action of the $j$-th generator, we impose the \emph{quasi-periodic
conditions} 
\begin{equation}
\psi(v_{+})=e^{i\alpha_{j}}\psi(v_{-}),\quad\psi'(v_{+})=-e^{i\alpha_{j}}\psi'(v_{-}).\label{eq:quasi_per_cond}
\end{equation}
We remind the reader that we use the convention of always taking the
derivatives into the edge, which explains the minus sign in conditions~(\ref{eq:quasi_per_cond}).
For example, in the fundamental cell of Fig.~\ref{fig:per_1d}(b)
the empty circles denote the vertices connected through the condition
of the above type. Identifying these periodically related vertices
creates new cycles, $C_{j}$, $j=1,\ldots,J$, on the graph and the
resulting problem is equivalent to a graph with magnetic fluxes $\alpha_{j}$
through the corresponding cycles. 
For example, the result of the Floquet--Bloch procedure for the fundamental
cell in Fig.~\ref{fig:per_1d}(d) is equivalent to the \emph{magnetic}
graph in Fig.~\ref{fig:loop_and_dangling_graph}(a). We denote by
$E$ the number of edges of the resulting magnetic graph.

Expanding the solutions to \eqref{eq:Hamiltonian} in the basis of
$e^{\pm ikx}$ and applying the vertex conditions leads, after some
linear algebra (see \citep{KotSmi_prl97}), to the secular equation
\begin{equation}
F(k;\vec{\alpha}):=\det\left(\id-\rme^{i\left(\bA+k\bL\right)}\bS\right)=0,\label{eq:sec_eq}
\end{equation}
where all matrices act in the space of coefficients on directed edges;
each edge gives rise to two directed edges of equal length, therefore
all matrices have degree $2E$. The diagonal matrix $\bL$ is the
matrix of lengths of the directed edges. The diagonal matrix $\bA$
contains the magnetic fluxes $\alpha_{j}$ that are put upon the edges
created by vertex identifications. The magnetic fluxes change sign
when reversing the direction of the corresponding edge. Finally, the
unitary matrix $\bS$ contains directed edge-to-edge scattering coefficients,
which, for scattering at a Neumann-Kirchhoff vertex of degree $d$,
is equal to $-1+2/d$ for back-scattering and $2/d$ for forward scattering.
Most importantly, for our vertex conditions the matrix $\bS$ is independent
of $k$. See \eqref{eq:bA_example}--\eqref{eq:bS_example}, which
show these matrices for a specific graph.

Next we apply a clever trick originally due to Barra and Gaspard \citep{BarGas_jsp00}
(see also \citep{BerWin_tams10}): we introduce a new function $\Phi(\vec{\kappa};\vec{\alpha})$
such that 
\begin{equation}
\Phi(\kappa_{1}=kl_{1},\ldots,\kappa_{E}=kl_{E};\vec{\alpha}):=F(k;\vec{\alpha}),\label{eq:sec_fun_defined_on_torus}
\end{equation}
where $l_{1},\ldots,l_{E}$ are the graph edge lengths. A cursory
look at equation~\eqref{eq:sec_eq} reveals that the variables $\kappa_{e}$,
$e=1,\ldots,E$ need only be known modulo $2\pi$. For a fixed $\vec{\alpha}$,
define $\Sigma_{\vec{\alpha}}$ to be the set of solutions of 
\begin{equation}
\Phi(\vec{\kappa};\vec{\alpha})=0,\label{eq:Sigma_alpha}
\end{equation}
on the torus $\T_{E}:=\left[0,2\pi\right)^{E}$. Then the roots $k_{n}$
of the equation $F(k;\vec{\alpha})=0$ can be interpreted as the times
($k$ values) of piercing of the set $\Sigma_{\vec{\alpha}}$ by the
flow 
\begin{equation}
\vec{\kappa}\left(k\right)=k\cdot(l_{1},l_{2},\ldots,l_{E})\mod2\pi.\label{eq:flow}
\end{equation}
We now conclude that $k$ belongs to the spectrum, $\sigma$, of the
periodic graph if the corresponding point $\vec{\kappa}(k)$ belongs
to the set $\Sigma_{\vec{\alpha}}$ for \emph{some} value of $\vec{\alpha}$
(which itself belongs to a $J$-dimensional torus). For future purposes
we define 
\begin{equation}
\Sigma=\bigcup_{\vec{\alpha}\in[0,2\pi)^{J}}\Sigma_{\vec{\alpha}}.\label{eq:all_Sigma}
\end{equation}
We will now compute the set $\Sigma$ in a simple but important example
and then proceed to discuss how the questions about the band probability
$\ps$ can be related to the properties of the set $\Sigma$.


\paragraph*{\uline{Loop with an edge.}}

We now compute the set $\Sigma$ for a graph which consists of a loop
pierced by magnetic field with flux $\alpha$ and a single edge attached,
see Fig.~\ref{fig:loop_and_dangling_graph}(b). 

The numbering of the directed edges is given in Fig.~\ref{fig:loop_and_dangling_graph}(b).
According to this numbering the matrices $\bA$, $\bL$ and $\bS$
are given by 
\begin{equation}
\bA=\diag(\alpha,0,-\alpha,0),\quad\bL=\diag(l_{1},l_{2},l_{1},l_{2}).\label{eq:bA_example}
\end{equation}
and 
\begin{equation}
\bS=\left(\begin{smallmatrix}\nicefrac{2}{3} & \nicefrac{2}{3} & -\nicefrac{1}{3} & 0\\
0 & 0 & 0 & 1\\
-\nicefrac{1}{3} & \nicefrac{2}{3} & \nicefrac{2}{3} & 0\\
\nicefrac{2}{3} & -\nicefrac{1}{3} & \nicefrac{2}{3} & 0
\end{smallmatrix}\right)\label{eq:bS_example}
\end{equation}

\begin{figure}[!]
\setlength{\unitlength}{\columnwidth} \begin{picture}(1,0.52) \put(0,0.07){\includegraphics[scale=0.16]{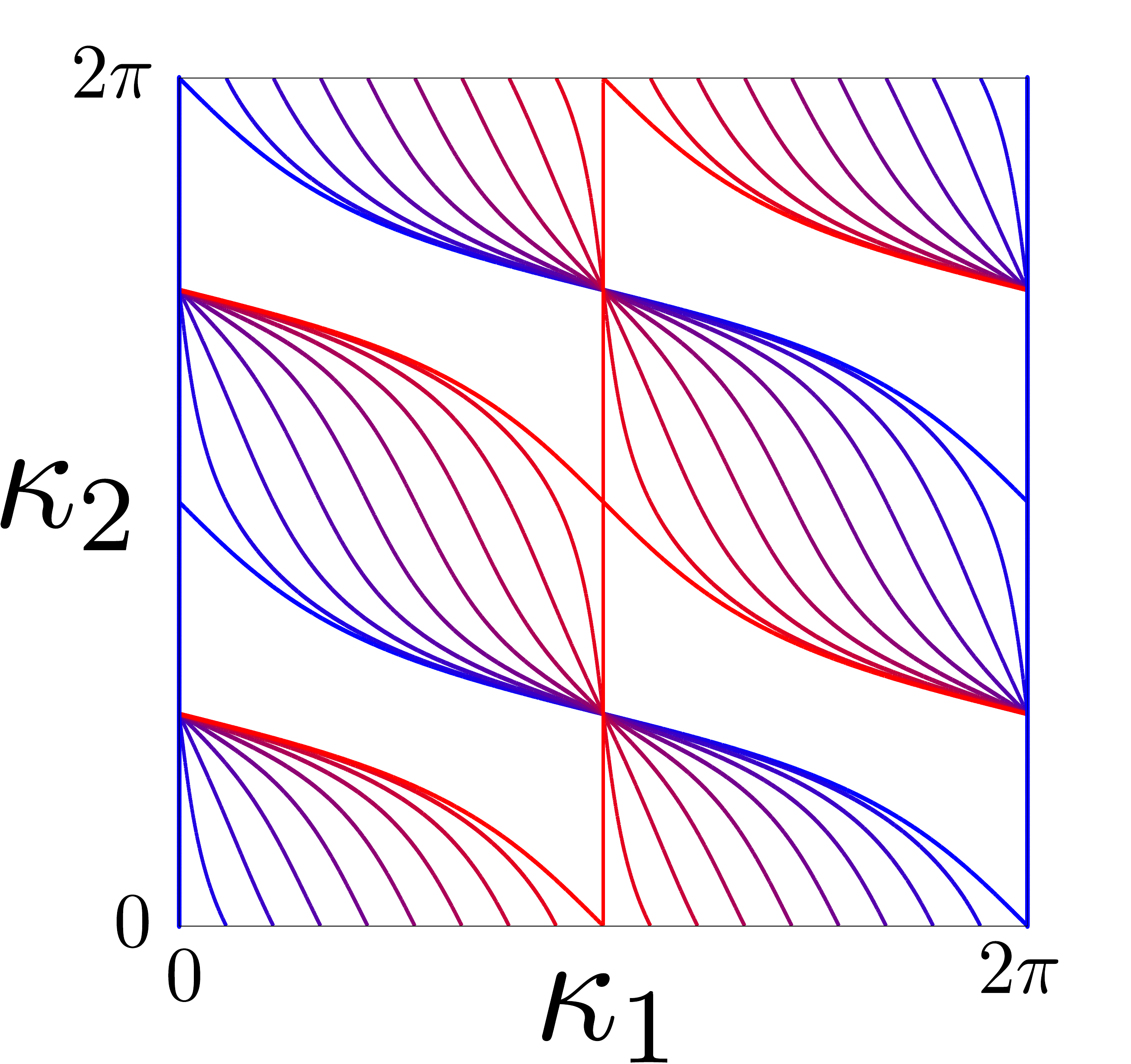}}\put(0.23,0){(a)}\put(0.5,0.07){\includegraphics[scale=0.16]{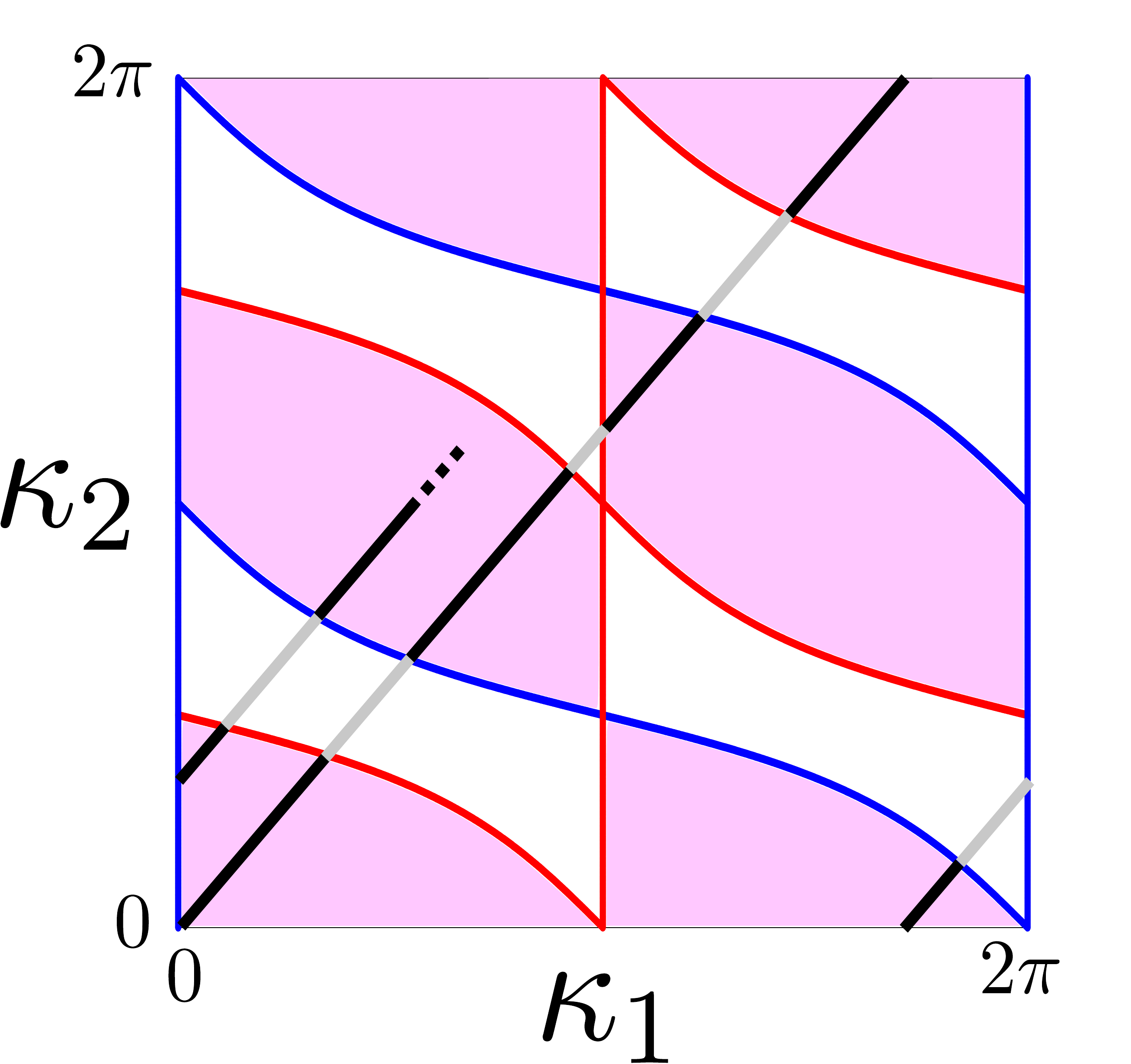}}\put(0.73,0){(b)}\end{picture}%
 \caption{(color online) (a) The zero sets, $\Sigma_{\alpha},$ of $\Phi\left(\kappa_{1},\kappa_{2};\,\alpha\right)$
are shown for a range of values $\alpha\in\left[0,\pi\right]$ using
a blue ($\alpha=0$) - red ($\alpha=\pi$) color scale. (b) The set
$\Sigma=\bigcup_{\alpha\in[0,2\pi)}\Sigma_{\alpha}$ is shaded, its
blue boundaries are the zeros of $\Phi\left(\cdot,\cdot;\,\alpha=0\right)$
and the red boundaries are the zeros of $\Phi\left(\cdot,\cdot;\,\alpha=\pi\right)$.
A flow $\vec{\kappa}\left(k\right)=k\cdot\left(l_{1},l_{2}\right)$
on the torus is indicated. The bands of the spectrum $\sigma$ are
the solid black segments of the flow line; the gaps are drawn in light
gray.}

\label{fig:torus} 
\end{figure}

The secular function $\Phi$ evaluates to (up to some non-zero factors)
\begin{multline}
\Phi=2\cos(\kappa_{2})\left(\cos(\kappa_{1})-\cos(\alpha)\right)\\
-\sin(\kappa_{1})\sin(\kappa_{2}).\label{eq:Phi_lasso}
\end{multline}
The zero sets $\Sigma_{\alpha}$ for a range of values of the parameter
$\alpha$ are shown on Fig.~\ref{fig:torus}(a). Note that it is
enough to consider the values $\alpha\in\left[0,\pi\right]$ as $\Sigma_{-\alpha}=\Sigma_{\alpha}$
(see \eqref{eq:Phi_lasso}).

\paragraph*{\uline{Probability to be in the spectrum.}}

From the discussion above we conclude that the probability $\ps$
for a random $k$ to be in the spectrum $\sigma$ is equal to the
proportion of time the flow defined by \eqref{eq:flow} spends in
the set $\Sigma$. Depending on the commensurability properties of
the set of the edge lengths, $\left\{ l_{e}\right\} _{e=1}^{E}$,
the flow covers densely the entire torus or is restricted to a flat
submanifold 
\begin{eqnarray*}
L & := & \overline{\spn\left(k\cdot(l_{1},l_{2},\ldots)\,\mod\T_{E}:k\in\R\right)}\\
 & = & \left\{ x\in\R^{E}:Mx=0\right\} \,\mod\T_{E},
\end{eqnarray*}
where $M$ is a matrix with rational coefficients (it gives the rational
dependencies in the length sequence $(l_{1},\ldots,l_{E})$). In the
latter case, the flow is ergodic\emph{ }on the submanifold $L$. The
probability $\ps$ is therefore the relative volume 
\begin{equation}
\ps=\frac{\vol_{L}\left(L\cap\Sigma\right)}{\vol_{L}(L)},\label{eq:volume_expr_for_p_s}
\end{equation}
where the subscript $L$ indicates that the volume should be taken
in the appropriate dimension (equal to $E$ minus the rank of the
matrix $M$). Formula \eqref{eq:volume_expr_for_p_s} remains valid
in the case of rationally independent lengths, when we simply take
$L$ to be the entire torus. This immediately implies that \emph{the
probability $p_{\sigma}$ remains the same as long as the edge lengths
are rationally independent}.

Returning to our example, we calculate $\ps$ explicitly. Using symmetry
we compute the area of $1/8$-th of the set $\Sigma$, the part in
the lower left corner. It is bounded by the coordinate axes and the
set $\Sigma_{\pi}$, which from \eqref{eq:Phi_lasso} we re-parameterize
as 
\begin{equation}
\tan(\kappa_{2})=2\cot(\kappa_{1}/2).\label{eq:Sigma_pi}
\end{equation}
Therefore the ratio in \eqref{eq:volume_expr_for_p_s} evaluates to
\begin{equation}
\ps=\frac{2}{\pi^{2}}\int_{0}^{\pi}\tan^{-1}\left(2\cot\left(\nicefrac{\kappa}{2}\right)\right)d\kappa\approx0.64.\label{eq:p_s_example}
\end{equation}

We can further prove that this universality of $p_{\sigma}$ extends
to a certain class of decoration structures. These are the decorations
that attach to the base line by means of a single edge, as in Fig.~\ref{fig:per_1d}(a)
to (d). Proving the universality is done by reducing the influence
of the decoration on the secular equation to a single scattering reflection
phase located at the degree one vertex of the graph in figure 
\ref{fig:loop_and_dangling_graph}(b). The phase enter the matrix
$\bS$ as follows, 
\begin{equation}
\bS=\left(\begin{smallmatrix}\nicefrac{2}{3} & \nicefrac{2}{3} & -\nicefrac{1}{3} & 0\\
0 & 0 & 0 & \Theta\left(\kappa_{3},\ldots,\kappa_{E}\right)\\
-\nicefrac{1}{3} & \nicefrac{2}{3} & \nicefrac{2}{3} & 0\\
\nicefrac{2}{3} & -\nicefrac{1}{3} & \nicefrac{2}{3} & 0
\end{smallmatrix}\right).\label{eq:bS_with_phase}
\end{equation}
While the precise form of the phase $\Theta\left(\kappa_{3},\ldots,\kappa_{E}\right)$
may be complicated, its effect on the function $\Phi$ gets averaged
out by ergodicity. More precisely, we now assume that the rational
relations (if any) defining the submanifold $L$ do not involve $\kappa_{1}$
and $\kappa_{2}$. In other words, the lengths of the edges $1$ and
$2$ are rationally independent of each other and of the lengths of
the decoration's edges. We need not assume anything about the lengths
of edges of the decoration.

One can now easily read from the determinant (see \eqref{eq:sec_eq},\eqref{eq:sec_fun_defined_on_torus})
that the function $\Phi$ has the form $\Phi\left(\kappa_{1},\kappa_{2},\ldots,\kappa_{E};\,\alpha\right)=\Phi\left(\kappa_{1},\kappa_{2}+\nicefrac{1}{2}\Theta\left(\kappa_{3},\ldots,\kappa_{E}\right);\,\alpha\right)$,
where $\Phi\left(\cdot,\cdot;\,\alpha\right)$ in the RHS is as in
\eqref{eq:Phi_lasso}. Introducing the change of variables 
\begin{equation}
\hat{\kappa}_{2}=\kappa_{2}+\frac{1}{2}\Theta\left(\kappa_{3},\ldots,\kappa_{E}\right),\label{eq:var_change}
\end{equation}
the integrals in \eqref{eq:volume_expr_for_p_s} factorize. Namely,
denote by $\T_{2}$ the torus with respect to $\kappa_{1}$ and $\hat{\kappa}_{2}$
and by $\T_{E-2}$ the torus with respect to the other variables.
Note that the set $\Sigma$ depends only on the variables $\kappa_{1}$
and $\hat{\kappa}_{2}$ (and is cylindrical with respect to the other
variables). The submanifold $L$, on the other hand, is cylindrical
with respect to $\kappa_{1}$ and $\hat{\kappa}_{2}$. Therefore 
\begin{equation}
\ps=\frac{\vol_{\T_{2}}\left(\T_{2}\cap\Sigma\right)\vol_{\T_{E-2}}\left(\T_{E-2}\cap L\right)}{\vol_{\T_{2}}\left(\T_{2}\right)\vol_{\T_{E-2}}\left(\T_{E-2}\cap L\right)},\label{eq:p_s_gen_decor}
\end{equation}
reducing to the expression in \eqref{eq:volume_expr_for_p_s}, where
$L$ there is identified as $\T_{2}$ in \eqref{eq:p_s_gen_decor}.
We thus proved that for all decorations of the type discussed above
the probability to be in the spectrum is given by \eqref{eq:p_s_example}.

To give a final example of a different nature, for the fundamental
cell depicted in Fig.~\ref{fig:per_1d}(e), the secular equation
can be shown to be equivalent to 
\begin{multline}
\sin(\kappa_{1}+\kappa_{2}+\kappa_{3})-\frac{1}{2}\sin\kappa_{1}\sin\kappa_{2}\sin\kappa_{3}\\
=\sin\kappa_{1}+\cos\alpha(\sin\kappa_{2}+\sin\kappa_{3}),\label{eq:sec_eq_dihedral}
\end{multline}
and the corresponding value of $p_{\sigma}$ was calculated numerically
to be $0.43$.


\paragraph*{\uline{Conclusions.}}

The arguments presented above apply to all graphs and result in three
general conclusions. First, given a $\Z^{d}$-periodic graph with
an arbitrary fundamental cell, the probability $\ps$ is independent
of the specific edge lengths, as long as there are no rational dependencies
between some of them. Even if such dependencies exist, an appropriate
ergodicity argument shows that the limit \eqref{eq:p_s_probability}
which defines $\ps$ exists and its exact value depends on the nature
of the edge lengths rational dependencies (as well as the graph's
topology). Secondly, we have shown that $\ps$ is robust even within
some topological modifications of the graph - attaching a prescribed
class of decorations. Thirdly, if there exists at least one non-flat
band (resp.\ gap) in the spectrum, it must arise from an open set
on the torus which is a subset of $\Sigma$ (corresp.\ $\T\backslash\Sigma$).
The ergodic flow on the torus will pass through this set infinitely
many times, resulting in an infinite number of non-flat bands (resp.\ gaps)
of comparable size. From equation~(\ref{eq:volume_expr_for_p_s})
we can immediately conclude that $\ps>0$ (resp.\ $\ps<1$).

Our setup calls for comparison with periodic potentials on the line,
in particular the singular potentials $\delta$ and $\delta'$ \citep{AlbGes+_solvable}.
Note that we measure our band and gap sizes in terms of the momentum
variable $k$, not energy (which scales as $k^{2}$). For smooth periodic
potentials and $\delta$ potentials, the gaps sizes decrease as $k\to\infty$,
while the band lengths converge to a constant, resulting in $\ps=1$
\citep{WeiKel_siamjam87,ExnGaw_prb96}. The $\delta'$ potential has
an opposite behavior, asymptotically equivalent to disconnecting the
graph: the band lengths decrease and the gaps approach a constant
size, resulting in $\ps=0$ \citep{ExnGaw_prb96}. Our results show
that a typical non-trivial periodic graph has intermediate behavior
with $0<\ps<1$, as long as there is at least one gap and at least
one band. One explanation of this phenomenon is that the graph of
Fig.~\ref{fig:per_1d}(a) (for example) can be viewed as a line with
periodic $\delta$-potential (the so-called Kronig-Penney model) whose
strength is momentum dependent. In such an analogue, the strength
of the $\delta$-potential oscillates between infinity and zero, which
in effect alternates between disconnecting the graph and having a
perfect transmission, resulting with the intermediate values $0<\ps<1$.
We refer the reader to \citep{AvrExnLas_prl94,ExnTur_jpa10} for similar
discussions.

One can also consider dressing the network with a bounded periodic
potential and/or changing the vertex conditions from the ones we considered.
This should not affect our results qualitatively, as the influence
of a potential or vertex conditions decreases in the $k\to\infty$
limit. However, this case is technically more difficult since the
$k$-dependence in equation (\ref{eq:sec_eq}) would become more involved.
To overcome these difficulties, methods developed in \citep{BolEnd_ahp09,RueSmi_jpa12}
might prove useful.

Some further interesting spectral questions are now within reach.
One may obtain bounds on possible sizes of bands (gaps) and deduce
the specific edge lengths for which they are attained. Furthermore,
the gap opening mechanism, a well studied subject on its own right
\citep{SchAiz_lmp00,Kuc_jpa05}, can be better understood by examining
the sub-domains of the torus which do not intersect $\Sigma$. In
addition, the topological meaning of $\ps$ should be further investigated
--- does it relate to some other graph invariants or does it provide
a brand new piece of information on the underlying graph?

Finally, we make another step forward by extending the discussion
to eigenfunction properties. The number of zeros of an eigenfunction
was recently found to be connected with the stability of the corresponding
eigenvalue with respect to magnetic perturbations \citep{Ber_prep11,Ver_prep12,BerWey_ptrs13}.
The stability is described by the Morse index of the eigenvalue and
most strikingly, this Morse index can be shown to be a well defined
function on the torus, not depending on the direction of the flow
(i.e., on graph edge lengths) \citep{Ban_ptrs13}. This leads to new
and exciting findings on the distribution of number of zeros of graph
eigenfunctions \citep{BanBer_surplus_distribution}.


\paragraph*{Acknowledgments.}

We thank D. Cohen, P. Exner and P. Kuchment for interesting discussions
and helpful advice. RB acknowledges the support of EPSRC, grant number
EP/H028803/1. GB acknowledges the support of NSF grant DMS-0907968.
The collaboration between the authors benefited from the support of
the EPSRC research network \textquoteleft{}\textquoteleft{}Analysis
on Graphs\textquoteright{}\textquoteright{} (EP/I038217/1).

\bibliographystyle{apsrev}
\bibliography{Band_Gap_Ratio}

\end{document}